\documentclass[12pt]{article}
\usepackage{latexsym}
\usepackage{amssymb}
\usepackage{amsmath}
\usepackage{amsfonts}
\usepackage{amssymb}
\usepackage{amsthm}
\usepackage{indentfirst}
\usepackage{mathrsfs}
\usepackage{graphicx}
\usepackage{color}

\catcode`@=11
\renewcommand{\section}{\@startsection{section}{1}{0pt}{\medskipamount}
{\medskipamount}{\large\bf}} \numberwithin{equation}{section}
\catcode`@=12

\def\sfrac#1#2{{\textstyle\frac{#1}{#2}}}

\def\a{{\alpha}}

\def\b{{\beta}}

\def\d{{\delta}}

\def\f{\frac}

\def\la{\label}
\def\eq{\eqref}
\def\pr{\partial}

\newcommand{\be}{\begin{equation}}
\newcommand{\ee}{\end{equation}}
\def\tr{{\rm tr}}
\newcommand{\Tr}{{\rm Tr}}

\def\bea{\begin{eqnarray}}
\def\eea{\end{eqnarray}}

\def\pa{\partial}                       

\def\beq{\begin{eqnarray}}    
\def\eeq{\end{eqnarray}}      

\def\ln{\,\mbox{ln}\,}                  
\def\tr{\,\mbox{tr}\,}                  
\def\Tr{\,\mbox{Tr}\,}                  
\def\det{\,\mbox{det}\,}                
\def\Det{\,\mbox{Det}\,}                

\begin{document}

\begin{center}

\phantom{x}

\vspace{1cm}

{\Large\bf Legendre transformations and   Clairaut-type  equations}

\vspace{18mm}

{\bf Peter M. Lavrov$^{(a, b)}\footnote{E-mail:
lavrov@tspu.edu.ru}$,\; Boris S. Merzlikin$^{(c)}\footnote{E-mail:
merzlikin@tspu.edu.ru}$\;}

\vspace{5mm}

\noindent  ${{}^{(a)}} ${\em Tomsk State Pedagogical University,
 Kievskaya St.\ 60, 634061 Tomsk, Russia}

\noindent  ${{}^{(b)}} ${\em National Research Tomsk State
University, Lenin Av.\ 36, 634050 Tomsk, Russia}

\noindent  ${{}^{(c)}} ${\em National Research Tomsk Polytechnic
University, Lenin Av.\ 30, 634050 Tomsk, Russia}

\vspace{20mm}

\begin{abstract}
\noindent
It is noted that the Legendre transformations in
the standard formulation of quantum field theory have the form of
functional Clairaut-type equations. It is shown that in presence of composite
fields the Clairaut-type form holds after loop corrections are taken into account.
A new solution to the functional Clairaut-type equation
appearing in field theories with composite fields is found.
\end{abstract}

\end{center}

\vfill

\\


\newpage

\section{Introduction}

In quantum field theory the main object containing all possible
information about a given dynamical system in quantum field theory
is the generating functional of vertex functions or, in other words,
the effective action. The usual way to introduce the effective
action is by means of the Legendre transformation applied to the
generating functional of connected Green functions.  The relation
between the effective action and the generating functional of
connected Green functions has the form of functional Clairaut-type
equation (see recent discussion in \cite{WD}). However, the
perturbative (loop) expansion in the effective action does not
preserve the mentioned Clairaut-type form of the Legendre
transformation. This may explain why in the quantum field theory
this specific feature of the Legendre transformation has been
ignored.

In the present paper we are motivated by recent works \cite{LS,LM}
devoted to the development of a new method to the functional
renormalization group approach \cite{P,W,W2} and the study of the
average effective action with composite fields. The approach to
quantum field theory with composite fields has been developed by
Cornwall, Jackiw and Tomboulis \cite{CJT} in attempts to study
physical phenomena (spontaneous symmetry breakdown, bound states,
etc.) which cannot be easily considered in the perturbation (loop)
expansion. Generalization of this method to gauge theories \cite{L}
detected a special form of gauge dependence of the effective action
with composite fields which in turn allowed to formulate the
approach to functional renormalization group \cite{LS} being free of
the gauge dependence problem inherent to the standard one
\cite{W,W2}. Introduction of the effective action with composite
fields requires to use the double Legendre transformations which as
compared with the standard case cannot be presented in the form of
functional Clairaut-type equation. Nevertheless, the perturbation
series   in the effective action with composite fields leads exactly
to the functional Clairaut-type equation.

The paper is organized as follows. In Section 2 the relations
existing between the effective actions without composite fields and
with composite fields and functional Clairaut-type equations are
derived. In Section 3 we study in detail solutions to the
first-order partial Clairaut-type equations with a special form of
the right-hand side and then generalize this result to the case of
functional Clairaut-type equations. In Section 4 the way to
find the one-loop correction to the effective action with composite
fields by solving an appropriate functional Clairaut-type
equation is shown. The remarkable result is that the effective action without
composite fields does not offer such a solution. In Appendix A the
simplest example of a set of matrices playing a very important
role in solving the functional Clairaut-type equation which appears
in field theories with composite fields is given.
\\

\section{Effective actions and Clairaut-type equations}

Let us consider a field model which is described by
a non-degenerate  action, $S[\phi]$, of the scalar field $\phi=\phi(x)$.
The generating functional of the
Green functions, $Z[J]$, and the generating functional of the
connected Green functions, $W[J]$, are defined in the standard
way\footnote{We use the DeWitt's condensed notations \cite{DeWitt} and
restrict ourselves to the case of a real scalar field for the sake of simplicity.}
 \bea
Z[J] = \int {\cal D}\phi\; e^{i \left(S[\phi]+J\phi\right)}=
e^{iW[J]}\,, \la{Z1}
 \eea
where $J=J(x)$ is usual source to $\phi$ and $J\phi=\int dx J(x)\phi(x)$.
The effective action,
$\Gamma=\Gamma[\Phi]$, is introduced by the Legendre transformation
of $W[J]$
 \bea
\Gamma[\Phi] = W[J] - J \Phi\,, \la{ea1}
\eea
\bea
 \f{\d W[J]}{\d J(x)} = \Phi(x)\,, \qquad
 \f{\d \Gamma[\Phi]}{\d \Phi(x)} = - J(x)\,. \la{ea2}
 \eea
Eliminating the source $J$ form \eq{ea1} one obtains the  equation
($\Gamma=\Gamma[\Phi]$)
 \bea
 \Gamma- \f{\d \Gamma}{\d \Phi} \Phi
 = W\left[-\f{\d \Gamma}{\d \Phi}\right]\,, \la{Cleq1}
 \eea
which has exactly the form of Clairaut-type equation. The practical
application of Eq. (\ref{Cleq1}) requires solving the equation
\eq{Z1} for $W[J]$. The standard way to find a solution to Eq.
(\ref{Cleq1}) with the functional $W$ defined in Eq. (\ref{Z1}) is
related to using the perturbation theory. In the tree approximation,
$\Gamma[\Phi]=\Gamma^{(0)}[\Phi]$, it follows
$\Gamma^{(0)}[\Phi]=S[\Phi]$. In the one-loop approximation,
$\Gamma[\Phi]=S[\Phi]+\Gamma^{(1)}[\Phi]$, one derives \bea
\label{G1} \Gamma^{(1)}[\Phi]=\frac{i}{2}\Tr\ln S^{''}[\Phi],
\eeq
where $S^{''}[\Phi]$ means the functional matrix with elements
\bea
S^{''}[\Phi](x,y)=\frac{\delta^2 S[\Phi]}{\delta
\Phi(x)\delta\Phi(y)}\;,
\eea
and so on. Here and below we assume that the functional trace, $\Tr$,
can be defined in an appropriate way for functional matrices under consideration.
In what follows we will use the following properties of functional trace
\beq
\Tr (M+N)=\Tr M +\Tr N, \quad \Tr MN=\Tr NM\,,
\eeq
for any two suitable functional matrices $M$ and $N$ as well as the relation between
the functional determinant and the functional trace
\beq
\Det M=\exp\{\Tr \ln M\}\,.
\eeq

In fact, specific features of Eq. (\ref{Cleq1}) being
the Clairaut-type equation disappear within the perturbation expansion.
But it is not the case when
one studies the effective action with composite fields \cite{CJT}.
Indeed, let us consider the effective action with composite
fields. The starting point of such an approach is the generating
functional of Green functions $Z[J, K]$,
 \bea
 Z[J,K] = \int {\cal D} \phi\; e^{i \left(S[\phi]+J\phi +
  K L(\phi)\right)} = e^{iW[J,K]}\,, \la{Z2}
 \eea
where $W=W[J,K]$ is the generating functional of connected Green
functions. In Eq. (\ref{Z2}) $J=J(x)$ and $K=K(x,y)$ are sources to field
$\phi=\phi(x)$ and composite field $L(\phi)=L(\phi)(x,y)$, respectively,   and
the notation $K L(\phi)=\int dx dy\, K(x,y)L(\phi)(x,y)$ is used.
Let  $L(\phi)(x,y)$ depend quadratically on the field $\phi$
\beq
L(\phi)(x,y)=\f12 \phi(x)\phi(y)\;.
\eeq
The effective action with composite field,
$\Gamma=\Gamma[\Phi,F]$, is defined by using the double Legendre
transformation \cite{CJT}
 \bea
 \Gamma[\Phi, F] = W[J, K] - J \Phi - K
 \big(L(\Phi) + \sfrac 12 F\big)\,, \la{ea3}
 \eea
 \bea
\f{\d W[J,K]}{\d J(x)} = \Phi(x)\,, \qquad  \f{\d W[J,K]}{\d
K(x,y)} = L(\Phi)(x,y) + \sfrac12 F(x,y) \,, \la{ea4} 
\eea
\bea
\f{\d \Gamma[\Phi, F]}{\d \Phi(x)} = - J(x) - \int dy
K(x,y)\Phi(y)\,, \qquad \f{\d \Gamma[\Phi, F]}{\d F(x,y)} = -
\sfrac12 K(x,y)\,. \la{ea5}
 \eea
Eliminating the sources $J$ and $K$ from the \eq{ea3} one
obtains the equation
 \bea
 \Gamma- \f{\d \Gamma}{\d \Phi} \Phi
 - \f{\d \Gamma}{\d F}F =
 W\left[-\frac{\delta\Gamma}{\delta\Phi}
 +  2 \frac{\delta\Gamma}{\delta F}\Phi, -2\frac{\delta\Gamma}{\delta
F}\right]
 - 2 \f{\delta\Gamma}{\d F} L(\Phi)\,. \la{Cleq2}
 \eea
Since the right-hand side of Eq. (\ref{Cleq2}) depends on the fields
$\Phi$ not only through derivatives of functional
$\Gamma=\Gamma[\Phi,F]$, the Eq. (\ref{Cleq2}) does not belong to
the Clairaut-type equation. But, in contrast with Eq. (\ref{G1}),  the
one-loop approximation for the effective action with composite
field, $\Gamma^{(1)}=\Gamma^{(1)}[\Phi, F]$ by itself satisfies
the
equation
 \bea
\Gamma^{(1)}- \f{\d \Gamma^{(1)}}{\d F} F
 = \f{i}{2} \Tr \ln\bigg(S^{''}[\Phi]
 -2 \f{\d \Gamma^{(1)}}{\d F}  \bigg)\,,\la{Cleq3}
 \eea
being exactly the Clairaut-type with respect to field $F$ wherein
the variable $\Phi$ should be considered as parameter.
\\

\section{Remarks on the solutions of Clairaut-type equations}

A Clairaut equation is a differential equation of the form
\beq
\label{eq1}
y - y' x = \psi (y'),
\eeq
where $y=y(x), y'=dy/dx$ and $\psi=\psi(z)$ is a real function of $z$. It is well-known that
the general solution of the Clairaut equation is the family of straight
line functions given by
\beq
\label{eq2}
y(x)=Cx+\psi(C),
\eeq
where $C$ is a real constant. The so-called singular solution is defined by the equation
\beq
\label{eq3}
\psi'(y')+x=0,
\eeq
if a solution to Eq. (\ref{eq3}) can be present in the form $y'=\varphi(x)$ with
a real function $\varphi$ of $x$. Then the solution to Eq. (\ref{eq3})
can be presented in the form
 \bea
 \label{eq4}
 y(x)=y(\xi)+\int_{\xi}^{x} dx\varphi (x).
 \eea

Now let us examine a first-order partial differential equation \cite{Kamke}
\bea
\label{e1}
 y - y'_i x^i = \psi (y'),
 \eea
which is also known as the Clairaut equation \cite{Kamke}. Here $y=y(x)$ is the real
function of variables $x\in \textbf{R}^n$, $\,x=\{x^1,x^2,...,x^n\},$
$\psi=\psi(z)$ is a real function of variables
$z=\{z_1,z_2,...,z_n\}$ and the notation
 \bea \label{e2}
y'_i = \pr_i y(x)\equiv \f{\pr y(x)}{\pr x^i}\,, 
 \eea
is used.  In terms of new function $z_i=z_i(x)=y'_i(x)$ the equation (\ref{e1})
rewrites
 \bea
\label{e3}
 y - z_i x^i = \psi (z).
 \eea
Differentiation of Eq. (\ref{e3}) with respect to $x^i$
leads to a system of differential equations
 \bea
 \label{e4}
 \frac{\pa z_j}{\pa x^i}\left(\frac{\pa \psi}{\pa z_j}+x^j\right)=0,\quad i=1,2,...,n.
 \eea
In the case of  the Hessian matrix vanishing $H_{ij}=0$ where
 \bea
 \label{e5}
H_{ij}=\frac{\pa z_j}{\pa x^i}=\frac{\pa^2 y}{\pa x^i\pa x^j},
 \eea
 $z_i=C_i={\rm const}$. Therefore the solution to Eq.(\ref{e1}) is the family of linear functions
 \bea
 \label{e6}
y(x)=C_ix^i+\psi (C),\quad C=\{C_1,C_2,...,C_n\}.
 \eea
If $\det H_{ij}\neq 0$ then the equations (\ref{e4}) are reduced to the following system
 \bea
 \label{e7}
 \frac{\pa \psi}{\pa z_j}+x^j=0, \quad j=1,2,...,n.
 \eea
If there are any real solutions to the equations (\ref{e7})
 \bea
 \label{e8}
 z_j=\varphi_j (x), \quad j=1,2,...,n\,,
 \eea
then the solution to Eq. (\ref{e1}) is reduced to the system of partial first-order
differential equations
 \bea
 \label{e9}
 y^{'}_j(x)=\varphi_j (x),\quad j=1,2,...,n\,,
 \eea
resolved with respect to derivatives. If the conditions of integrability
\beq
\pa_i\varphi_j(x)=\pa_j\varphi_i(x)
\eeq
are fulfilled then in any simply connected domain $G \subset\textbf{R}^n$ the  system (\ref{e9})
is solvable and the solution to this system can be presented in the form
\beq
y(x)=y(\xi)+\int_{\xi}^{x} dx^i \varphi_i(x)\,,
\eeq
where the integration is performed along any rectifiable curve in $G$ having endpoints
$\xi=\{\xi_1,\xi_2,...,\xi_n\}$ and $x=\{x_1,x_2,...,x_n\}$ (see, for example, \cite{Kamke}).

Now taking into account the structure of Eq.(\ref{Cleq3}) we consider the  Eq.(\ref{e3})
with the following choice of function $\psi$
 \beq
\label{e10}
\psi(z)=\a \ln\Big(1-\b\,(z_i a^i) \Big),
 \eeq
where $\alpha, \b$ are  real nonzero parameters and $a=\{a^1,a^2,...,a^n\}$ forms
a $n$-tuple constant vector in a vector space $V$.
If the
Hessian matrix \eq{e5} $H_{ij}=\pr_i z_j = 0\,,\forall
i,j=1,\ldots,n$ then  $z_i = C_i = {\rm const}$ and the
solution to the Eqs. \eq{e1}, \eq{e10} reads
 \bea
y(x)&=& C_i x^i  + \a \ln\Big(1-\b\, C_i a^i \Big)\,. \la{sol}
 \eea
In case when $\det H_{ij}\neq 0 $ the equations \eq{e7} for
$z_i=z_i(x)$ have the form
 \bea
-   \f{\a\b\, }{1-\b\,( z_ia^i)}\, a^j+ x^j=0\,, \quad j=1,2,...,n. \la{f}
 \eea
Specific feature of this system is that the required structure $z_ia^i$ and $z_ix^i$
as functions of $x=\{x_1,x_2,...,x_n\}$ can be found algebraically.
Indeed, let us introduce the constant vector $b=\{b_1,b_2,...,b_n\}$ in the dual
vector space $V^*$ so that
$(b_ia^i)=1$. Multiplying the Eq. \eq{f} by $b_j$ and summing the results one
obtains
 \bea
 \Big(1-\b (z_ia^i)\Big)^{-1} = \sfrac1{\a\b}\, (b_i x^i) \,, \quad \Rightarrow\quad
 z_ia^i = \b^{-1} - \a (b_ix^i)^{-1}\,.\la{e13}
  \eea
In turn multiplying the Eq. \eq{f} by $z_j$ and summing the results
we have
 \bea
  -\f{\a\b\,(z_ja^j)}{1-\b\, (z_ia^i)}+ (z_jx^j)=0\,, \quad \Rightarrow\quad
  z_i x^i = \b^{-1}(b_ix^i)-\a\,,    \la{e14}
 \eea
 where the equation \eq{e13} is used.
Finally the solution to Eqs. \eq{e1}, \eq{e10} can be presented in  the form
 \bea
 y(x)&=&  \b^{-1}(b_i x^i) -\a \ln(b_ix^i)-\a + \a \ln \a\b\,. \la{sol2}
 \eea

The results obtained above for the partial differential equations
can be immediately extended to the case of functional Clairaut-type
equations. Let $\Gamma=\Gamma[F]$ be a functional of fields $F^m=F^m(x)\,, m=1,2,...,N$,
which are real integrable functions of  real variables $x\in {\bf R}^n$.  We use the notion
of {\it functional Clairaut-type equations} for the equations of the form
\bea
\label{e16}
 \Gamma-\f{\d \Gamma}{\d F^m}F^m =\Psi\left[\frac{\delta\Gamma}{\delta F}\right],
 \eea
where $\Psi=\Psi[Z]$ is a given real functional of real variables
$Z_m=Z_m(x)\,, m=1,2,...,N$.
In Eq.(\ref{e16}) the following notation
 \bea
\label{e18}
\f{\d \Gamma}{\d F^m}F^m = \int d^n x \f{\d
\Gamma}{\d F^m(x)}F^m(x)
 \eea
is used. The functional derivatives are defined by the rule
 \bea
 \label{e17}
 \f{\d F^m(x)}{\d F^k(y)} = \d^m_k\, \d(x-y)\,.
 \eea

We restrict ourselves to the following
functional $\Psi$
\bea
\label{e19}
\Psi[Z]=\a\ln\Big( 1-\b (Z_m A^m) \Big)\,,
\eeq
where $A^m=A^m(x)\,, m=1,2,...,N$ is the set of given
integrable functions of real variables $x$ and $\alpha,\b$ are  real nonzero  parameters.
Omitting  the case of a linear functional $\Gamma$
and repeating almost word for word the arguments made above we derive the non-trivial solution
to the Eqs.(\ref{e16}), (\ref{e19})
 \bea
 \label{e20}
\Gamma[F] = \b^{-1} (B_m F^m) - \a\ln (B_m F^m)-\a + \a \ln\a\b\,,
 \eea
where $B_m=B_m(x),\quad m=1,2,...,N$, are field variables satisfying the condition
$B_mA^m=\int dx B_m(x)A^m(x)=1$.
\\

\section{The effective action with  composite fields}

In this Section we generalize the results obtained in \cite{CJT} to
the case of a field model described by a set of scalar bosonic
fields $\phi^A(x)\,, A=1,..,N$, with a classical non-degenerate action $S[\phi]$.
Let $L^i(\phi)=L^i(\phi)(x,y), i=1,2,...,M $, be composite non-local fields,
 \beq
 \label{L}
L^i(\phi)(x,y)=\frac{1}{2}\;{\cal A}^i_{AB}\;\phi^A(x)\phi^B(y)\;,
 \eeq
where ${\cal A}^i_{AB}={\cal A}^i_{BA}$ are constants. The generating functional of
Green functions, $Z[J, K]$, is given by the following path integral
 \bea
 \label{ZJK}
 Z[J, K]= \int {\cal D} \phi e^{i \left(S[\phi]+J_A\phi^A
  + K_i L^i(\phi) \right)} = e^{iW[J,K]}\,,
 \eea
where $K_i=K_i(x,y),\, i=1,2,...,M $, are sources to composite fields $L^i(\phi)(x,y)$.
Here the notations
 \bea
 J_A\phi^A &=& \int  dx\, J_A(x) \phi^A(x)\,, \\
 K_i L^i(\phi) &=& \int  dx dy\, K_i(x,y)L^i(\phi)(x,y)\,,
 \eea
are used.
From Eq. (\ref{ZJK}) we can construct the following relations
\beq
\frac{1}{2}{\cal A}^i_{AB}\frac{\delta^2 Z[J,K]}{\delta J_A(x)\delta J_B(y)}=
i\frac{\delta Z[J,K]}{\delta K_i(x,y)}\,,
\eeq
or, in terms of the functional $W[J,K]$,
\beq
\label{Wr}
\frac{1}{2}{\cal A}^i_{AB}\left[-i\frac{\delta^2 W[J,K]}{\delta J_A(x)\delta J_B(y)}
+\frac{\delta W[J,K]}{\delta J_A(x)}\frac{\delta W[J,K]}{\delta J_B(y)}\right]=
\frac{\delta W[J,K]}{\delta K_i(x,y)}\,.
\eeq

We define the average fields $\Phi^A(x)$ and composite fields
$F^i(x,y)$ as follows
 \bea
 \label{DL1}
 \f{\d W[J,K]}{\d J_A(x)} = \Phi^A(x)\,, 
 \eea
 \bea
 \label{DL2}
  \f{\d W[J,K]}{\d K_i(x,y)} &=& L^i(\Phi)(x,y) + \sfrac12 F^i(x,y)\,.
 \eea
 The effective action with composite fields,
$\Gamma=\Gamma[\Phi,F]$, is defined by using the double Legendre
transformation  of $W[J,K]$, (\ref{DL1}) and (\ref{DL2}),
 \bea
 \Gamma[\Phi, F] &=& W[J,K] - J_A \Phi^A
 - K_i\big( L^i(\phi) + \sfrac12F^i \big)\,. \la{ea4}
 \eea
One can eliminate the sources from Eq. \eq{ea4} using
 \bea
 \f{\d\Gamma[\Phi,F]}{\d \Phi^A(x)} = - J_A(x)
 - \int dy\, K_i(x,y) {\cal A}^i_{AB} \Phi^B(y) \,, 
 \eea
 \bea
 \f{\d\Gamma[\Phi,F]}{\d F^i(x,y)} = -\sfrac12 K_i(x,y)\,.
 \eea
The relation (\ref{Wr}) rewritten in terms of $\Gamma[\Phi, F]$
reads
 \beq
 \label{eqF}
F^j(x,y)-i(G^{-1})^{AB}(x,y){\cal A}^j_{AB}=0\,,
 \eeq
where $(G^{-1})$ is the matrix inverse to $G$,
 \beq
G=\{G_{AB}(x,y)\},\quad G_{AB}(x,y)=\Gamma^{''}_{AB}[\Phi,F](x,y)
-2\frac{\delta\Gamma[\Phi,F]}{\delta F^i(x,y)}{\cal A}^i_{AB}\,,
 \eeq
 and we have used the notation
 \beq
\Gamma^{''}_{AB}[\Phi,F](x,y)=\frac{\delta^2\Gamma[\Phi,F]}{\delta\Phi^A(x)\delta\Phi^B(y)}\,.
 \eeq

In the one-loop approximation, $\Gamma[\Phi, F] = S[\Phi] +
\Gamma^{(1)}[\Phi, F]$,  the equation for one-loop contribution,
$\Gamma^{(1)}$, to the effective action can be found using procedure
similar to \cite{LM}.  It has the form
 \bea
 \Gamma^{(1)} - \f{\d\Gamma^{(1)}}{\d F^i} F^i
 = \f{i}{2} \Tr \ln\Big(
S''_{AB}[\Phi] - 2 \frac{\delta\Gamma^{(1)}}{\delta F^i}{\cal
A}^i_{AB}\Big)\,,
 \la{e5.7}
 \eea
being  the exact functional Clairaut-type equation.

According to the general scheme which is discussed in the previous
section, to solve the Eq. \eq{e5.7} we introduce new functions
$Z_i(x,y)$
 \bea
 \frac{\delta\Gamma[\Phi,F]}{\delta F^i(x,y)} = Z_i(x,y)\,, \la{Zab}
 \eea
and substituting them to the Eq. \eq{e5.7} we obtain
 \bea
\Gamma^{(1)} &=& Z_i F^i
 + \f{i}{2} \Tr \ln Q\,,
 \la{GammaZ}
 \eea
where the matrix $Q=\{Q_{AB}(x,y)\}$ is defined as
\beq
\label{Q}
Q_{AB}(x,y)=S''_{AB}[\Phi](x,y) - 2 Z_i(x,y) {\cal A}^i_{AB}.
\eeq
Then varying the functional \eq{GammaZ} with respect to $F^i$
we obtain
\beq
\label{dQ}
\delta \Gamma^{(1)}=\delta Z_iF^i+Z_i\delta F^i+\frac{i}{2}\Tr Q^{-1}\delta Q,
\eeq
where $Q^{-1}$ is the inverse to $Q$
\beq
\int dz (Q^{-1})^{AC}(x,z)Q_{CB}(z,y)=\delta^A_B\;\delta(x-y).
\eeq
Taking into account the explicit form of $Q$ (\ref{Q}) and keeping in mind Eq. \eq{Zab}
from Eq. (\ref{dQ}) it follows
\bea
\nonumber
 \frac{\delta\Gamma^{(1)}[\Phi,F]}{\delta F^j(x,y)}
 &=& Z_j(x,y) + \int dz dz' \f{\d Z_i(z,z')}{\d F^j(x,y)} F^i(z,z')
 -\\
 &&-i \int dz dz' (Q^{-1})^{AB}(z,z')
 \f{\d Z_i(z',z)}{\d F^j(x,y)} {\cal A}^i_{BA} 
  = Z_j(x,y)\,.
 \eea
Thus the equation defining non-trivial functions $Z_i(x,y)$ reads
\bea
  F^j(x,y) - i  (Q^{-1})^{AB}(y,x){\cal A}^j_{BA} = 0\,.
  \la{eqPsi}
\eea
Note that in the approximation considered here
this equation coincides  with (\ref{eqF}).

To solve the equation (\ref{eqPsi}) we introduce a set of matrices
${\cal B}_{j}=\{{\cal B}_j^{AB}\}$ by the relations
\beq
\label{cB}
{\cal A}^j_{AB}{\cal B}_j^{CD}=
\frac{1}{2}\left(\delta^C_A\delta^D_B+\delta^D_A\delta^C_B\right).
\eeq
Then we have
\beq
F^j(x,y){\cal B}_j^{AB}-i(Q^{-1})^{BA}(y,x)=0,
\eeq
or, due to the symmetry property of $Q_{AB}(x,y)=Q_{BA}(y,x)$,
\beq
\label{FZ}
F^j(x,y){\cal B}_j^{AB}-i(Q^{-1})^{AB}(x,y)=0.
\eeq
From Eqs. (\ref{eqPsi}) and (\ref{FZ}) one deduces the relations
\beq
\label{res}
{\cal B}_j^{AB}{\cal A}^i_{BA}=\delta^i_j\quad {\rm or}\quad
\tr {\cal B}_j{\cal A}^i=\delta^i_j\,.
\eeq
In turn Eqs. (\ref{cB}) and (\ref{res}) lead to the restriction on parameters
$N$ and $M$
\beq
\label{cMN}
\frac{1}{2}N(N+1)=M\,.
\eeq
The condition (\ref{cMN}) has a simple sense: in a given theory with the set
of $N$ fields $\phi^A$
there exists exactly the $(1/2)N(N+1)$ independent combinations of $\phi^A\phi^B$.

The solution to the equation (\ref{FZ}) has the form
 \bea
 Z_i(x,y) {\cal A}^i_{AB} = \f{1}{2} S''_{AB}[\Phi](x,y)
 - \f{ i}{2}(F^i {\cal B}_i)^{-1}_{AB}(x,y)\,,
 \la{Zab2}
 \eea
where
\beq
\int dz (F^i {\cal B}_i)^{-1}_{AC}(x,z) (F^j(z,y) {\cal B}_j)^{CB}=\delta^A_B\delta (x-y)\,.
\eeq
We can express the $Z_i F^i$  as a functional of $\Phi,F$  multiplying
 Eq. \eq{eqPsi} by $Z_j$
and using  \eq{Zab2} with the result
 \bea
 \label{ZFf}
Z_jF^j = \f{1}{2} \Tr \Big((F^j {\cal B}_j) S''[\Phi]
\Big) -\f{ i}{2} \delta(0)N \,.
 \eea
Finally  substituting \eq{Zab2} into \eq{GammaZ} and using (\ref{ZFf}) we find the
one-loop effective action, $\Gamma^{(1)}[\Phi,F]$, in the form
 \bea
 \Gamma^{(1)}[\Phi, F] = \f{1}{2} \Tr \Big((F^j {\cal B}_j) S''[\Phi] \Big)
 -\f{i}{2} \Tr \ln \Big( i(F^j {\cal B}_j)\Big)
 -\f{ i}{2}\delta(0)N\,.
 \la{ea5}
 \eea
The expression for $\Gamma^{(1)}$ \eq{ea5} generalizes
the known result of \cite{CJT} in the cases $A=B=1$ and $j=1$ when ${\cal A}^1_{11}=1$ and
${\cal B}^{11}_1=1$.
This generalization involves the introduction of the set of matrices ${\cal B}_j$
obeying the properties (\ref{cB}). In Appendix A we give a simple example of matrices
${\cal A}^j$ and ${\cal B}_j$ satisfying all required properties.
\\

\section{Discussions}
In the present article we have studied relations existing between the Legendre transformations
in quantum field theory and the functional differential equation for effective action which has
the form of functional Clairaut-type equation. We have found that
specific features of this equation do not hold within the perturbation theory
in a quantum field theory without composite operators.
But it is not the case within the approach to the quantum field theory
based on composite fields when perturbation expansion of the effective action leads exactly to
a functional Clairaut-type equation with a special type of the right-hand side.
Partial first-order
differential equations of Clairaut-type were our preliminary step in the study of solutions
to the problem. It was shown that in case when the right-hand side of the equation
has the form inspired by the real situation in quantum field theory with composite fields
the solution to that functional Clairaut-type equation can be found
with the help of algebraic manipulations only. In our knowledge the solution (\ref{sol2})
to the equations (\ref{e1}) and (\ref{e10}) can be considered as a new result in the theory
of partial first-order differential equations of Clairaut-type. This result has been easily
extended (see (\ref{e20})) to the case of functional Clairaut-type equation (\ref{e16})
with the special right-hand side (\ref{e19}). We have found an explicit solution to the functional
Clairaut-type equation appearing in the quantum field theory with composite fields to define
one loop contribution to the corresponding effective action (\ref{ea5}).

We have studied the case with maximum number  of composite fields,
$M=\f12 N(N+1)$, being quadratic in the given scalar fields
$\phi^A\,, A=1,...,N$. In a similar manner one can consider the
situation when the number of composite fields is less the maximum
one, $L^i(\phi)(x,y) =\f12 {\cal A}^i_{ab}\phi^a(x)\phi^b(y)\,,
i=1,...,M < \f12 N(N+1)\,,a=1,...,n<N$. Now the matrix of second
derivatives of the classical action, $S''_{AB}[\Phi]$, should be
presented  in the block form
 \bea
 S''_{AB}[\Phi](x,y) = \left(
                    \begin{array}{c|c}
                      S''_{ab} & S''_{a\b} \\
                      \hline
                      S''_{\a b} & S''_{\a \b}\\
                    \end{array}
                  \right)\,,
 \eea
where $\quad a,b = 1,...,n$ and $ \a,\b = n+1, ..., N\,.$ The
equation for the one-loop contribution, $\Gamma^{(1)}[\Phi, F]$, to
effective action takes the form
 \bea
 \Gamma^{(1)} - \f{\d \Gamma^{(1)}}{\d F^i} F^i = \f{i}2 \Tr
 \ln\Big({\tilde S}''_{ab}- 2\f{\d \Gamma^{(1)}}{\d
 F^i}{\cal A}^i_{ab}\Big) + \f{i}2 \Tr \ln S''_{\a \b}\,, \label{EAeq}
 \eea
where
 \bea
{\tilde S}''_{ab} = S''_{ab} - S''_{a\a} (S''^{-1})^{\a\b}S''_{\b
b}\,.
 \eea
The solution to the Eq. \eq{EAeq} reads
 \bea
 \Gamma^{(1)}[\Phi, F] =  \f{1}{2} \Tr \Big((F^j {\cal B}_j)^{ab} {\tilde S}''_{bc}[\Phi] \Big)
 -\f{i}{2} \Tr \ln \Big( i(F^j {\cal B}_j)^{ab}\Big) + \f{i}2 \Tr \ln
 S''_{\a\b}[\Phi]  -\f{ i}{2}\delta(0)n\,.
 \eea
Here the matrixes ${\cal B}^{ab}_i$ are introduced in the same way
as in \eq{cB} but for the ${\cal A}^i_{ab}$ ones.

Extension of the results obtained above to gauge theories can be
easily performed in a way used in papers \cite{LS,LM} on the basis
of supermathematics \cite{Ber1,Ber2}. We are going to present such
kind of generalizations in our further studies.

\section*{Acknowledgments}
\noindent
We thank S.E. Konshtein, O. Lechtenfeld, I.L. Shapiro and I.V. Tyutin for useful discussions.
The authors are thankful to the grant of Russian Ministry of
Education and Science, project 2014/387/122 for their support. The work is
also partially supported  by the Presidential grant 88.2014.2 for
LRSS and DFG grant LE 838/12-2.
\\

\appendix
\section*{Appendix A. The simplest example of the matrices ${\cal A}^i_{AB}$
and ${\cal B}_i^{AB}$}
\setcounter{section}{1}
\renewcommand{\theequation}{\thesection.\arabic{equation}}
\setcounter{equation}{0}

In solving the equation (\ref{eqPsi}) the existence of matrices
${\cal B}_i^{AB}$ satisfying the properties (\ref{cB}) and
(\ref{res}) plays a crucial role. To support this assumption we
consider a simple example. Namely, let us consider the matrices
${\cal A}^i_{AB}$ for the case of two fields, $N=2$, thus $A,B =
1,2$ in the Eq. \eq{L}. According to the Eq. \eq{cMN} we have
$M=3$ and $i=1,2,3$. The simplest choice of matrices ${\cal A}^i=\{{\cal
A}^i_{AB}\}$ reads
 \bea
{\cal A}^1 = \left(
                    \begin{array}{cc}
                      1 & 0 \\
                      0 & 0 \\
                    \end{array}
                  \right)\,, \qquad
{\cal A}^2 = \left(
                    \begin{array}{cc}
                      0 & 0 \\
                      0 & 1 \\
                    \end{array}
                  \right)\,, \qquad
{\cal A}^3 = \f12\left(
                    \begin{array}{cc}
                      0 & 1 \\
                      1 & 0 \\
                    \end{array}
                  \right)\,,
 \eea
or in the condensed notation
 \bea
{\cal A}^i = \left(
                    \begin{array}{cc}
                      \d^i_1 & \f12\d^i_3 \\
                      \f12\d^i_3 & \d^i_2 \\
                    \end{array}
                  \right)\,.
 \eea
Then it is easy to construct the matrixes ${\cal B}_i=\{{\cal B}^{AB}_i\}$
 \bea
{\cal B}_i = \left(
                    \begin{array}{cc}
                      \d_i^1 & \d_i^3 \\
                      \d_i^3 & \d_i^2 \\
                    \end{array}
                  \right)\,,
 \eea
which satisfy the conditions \eq{cB} and \eq{res}.
\\


\begin{thebibliography}{11}

\bibitem{WD}
M. Walker, S. Duplij, {\it Cho-Duan-Ge decomposition of QCD in the
constraintless Clairaut-type formalism}, Phys. Rev. \textbf{D91}
(2015) 064022.

\bibitem{LS}
P. M. Lavrov, I. L. Shapiro, {\it On the Functional Renormalization
Group approach for Yang-Mills fields}, JHEP {\bf 1306} (2013) 086.

\bibitem{LM} P. M. Lavrov, B. S. Merzlikin, {\it Loop expansion of the
average effective action in the functional renormalization group
approach}, Phys. Rev. \textbf{D92} (2015) 085038.

\bibitem{P}
J. Polchinski, {\it Renormalization and effective lagrangians},
Nucl. Phys. {\bf B231} (1984) 269.

\bibitem{W}
C. Wetterich, {\it Averege action and the renormalization group
equations}, Nucl. Phys. {\bf B352} (1991) 529.

\bibitem{W2}
C. Wetterich, {\it Exact evolution equation for the effective
potential}, Phys. Lett. {\bf B301} (1993) 90.

\bibitem{CJT} J. M. Cornwell, R. Jackiw, E. Tomboulis,
{\it Effective action for composite operators}, Phys. Rev.
\textbf{D10} (1974) 2428.

\bibitem{L}
P. M. Lavrov, {\it Effective action for composite fields in gauge
theories}, Theor. Math. Phys. {\bf 82} (1990) 282.

\bibitem{DeWitt}
B. S. DeWitt ,
{\it Dynamical theory of groups and fields},
(Gordon and Breach, 1965).

\bibitem{Kamke}
E. Kamke,
{\it Differentialgleichungen, Loesungsmethoden Und Loesungen, II
Partielle Differentialgleichungen Erster Ordnung Fuer Eine Gesuchte Funktion},
(Leipzig, 1959)

\bibitem{Ber1}
F. A. Berezin,
{\it The method of second quantization},
(Academic Press, New York, 1966).

\bibitem{Ber2}
F. A. Berezin,
{\it Introduction to superanalysis},
Mathematical Physics and Applied Mathematics, n$^{\rm o}$ 9,
(Reidel, Dordrecht, 1987).

\end{thebibliography}
\end{document}